\documentstyle[manuscript,eqsecnum,aps,version2]{revtex}
\draft
\input epsf

\begin{document}

\title
\bf Spinodal Decomposition in High Temperature
Gauge Theories
\endtitle

\author{Travis~R.~Miller and Michael~C.~Ogilvie}
\instit
Department of Physics, Washington University, St. Louis, MO 63130
\endinstit
\medskip
\centerline{\today}

\abstract

After a rapid increase in temperature across the deconfinement temperature $%
T_{d}$, pure gauge theories exhibit unstable long wavelength fluctuations in
the approach to equilibrium. This phenomenon is analogous to spinodal
decomposition observed in condensed matter physics, and also
seen in models of disordered chiral condensate formation. At
high temperature, the unstable modes occur only in the range $0\leq k$ $\leq
k_{c}$, where $k_{c}$ is on the order of the Debye screening mass $m_D$.
Equilibration always occurs via spinodal decomposition for $SU(2)\,$at
temperatures $T>T_{d}$ and for $SU(3)$ for $T\gg T_{d}$. For $SU(3)$ at
temperatures $T\gtrsim T_{d}$, nucleation may replace spinodal decomposition
as the dominant equilibration mechanism.
Monte Carlo
simulations of $SU(2)$ lattice gauge theory exhibit the predicted phenomena.
The observed value of $k_c$ is in reasonable agreement with
a value predicted from previous lattice measurements of $m_D$.

\endabstract


\section{Introduction}

In a heavy ion collision, changes in local energy densities of $1\ GeV/fm^3$
and higher occur on a time scale less than $1\ fm/c$.
During the initial stages of the formation of a quark-gluon
plasma, the system is thermodynamically unstable.
Some consequences of this instability can be studied
with both simulation and analytical techniques.
In particular, we find the existence of exponentially growing
long wavelength modes in the approach to equilibrium.
Such behavior is referred to as spindodal decomposition
in condensed matter physics.\cite{GuntonDroz}\cite{ChaikinLubensky}
We will study spinodal decomposition in finite temperature gauge theories
using both analytical and simulation techniques.
The existence of
exponentially growing long wavelength modes cutoff
at a fixed wavelength depends only on the features of the
equilibrium effective action in an unstable region.
Although the methods we use are not based on the true real-time evolution
of the system, they give a
consistent picture of some basic features of real-time behavior.

We consider first the generic case of a pure $SU(N)$
gauge theory in which the temperature is
raised rapidly from a temperature less than $T_{d}$, the
deconfinement temperature, to a temperature above $T_{d}$. 
We refer to such a rapid increase
in system temperature
as a quench; this nomenclature is borrowed from statistical
mechanics, where it is usually applied to rapid cooling of a system below a
critical temperature.
It is appropriate here because it is the high
temperature phase of the pure $SU(N)$ gauge theory which spontaneously
breaks global $Z(N)$ invariance.
At temperatures $%
T<T_{d}$, the $Z(N)$ global symmetry associated with confinement is
unbroken. The standard equilibrium order parameter for $Z(N)$ symmetry
breaking is the Polyakov loop $L$, defined in equilibrium as the
path-ordered exponential 
\begin{equation}
L(\overrightarrow{x})=\mathit{P}\exp \left[ i\int_{0}^{1/T}dt\,A_{0}(%
\overrightarrow{x},t)\right] 
\end{equation}
The Polyakov loop can be defined for a general density matrix by its
association with the projection operator onto gauge invariant states.
\cite{BanksUkawa}
Because of $Z(N)\,$symmetry, $%
\left\langle L_{F}\vspace{1pt}\right\rangle \,=\frac{1}{N}\left\langle
Tr_{F}(L)\right\rangle $ vanishes below $T_{d}$.
When the temperature is rapidly increased
to $ T > T_{d}$, $\left\langle L_{F}\vspace{1pt}\right\rangle \,=0$ is no
longer the stable state of the system, and must evolve to a new equilibrium
state with $\left\langle L_{F}\vspace{1pt}\right\rangle \,\neq 0$,
reflecting the transition to the gluon plasma phase. We will use the
Polyakov loop as the primary tool for the study of this process.

\section{Effective Potential}

The instability of $\left\langle L_{F}\vspace{1pt}\right\rangle \,=0$ at
high temperatures follows from the one-loop finite temperature effective
potential for the Polyakov loop, as derived by Gross, Pisarski and Yaffe and
Weiss.\cite{Gross}\cite{Weiss} For simplicity, consider the
case of $SU(2)$. At any spacetime point, the Polyakov loop can be
diagonalized, and we can write 
\begin{equation}
L=\left( 
\begin{array}{ll}
e^{i\pi q} &  \\ 
& e^{-i\pi q}
\end{array}
\right) \text{.} 
\end{equation}
where $0\leq q\leq 1$. It is convenient to introduce the parameter $\psi $,
which is related to $q$ by $q=\left( 1-\psi \right) /2$. The
order parameter $L_{F}$ is given by

\begin{equation}
L_{F}=\frac{1}{2}Tr_{F}\,L=\cos \left[ \pi \left( 1-\psi \right) /2\right] 
\text{.} 
\end{equation}
The effective potential at one loop for gauge bosons in a constant Polyakov
loop background is

\begin{equation}
V(\psi )=-\frac{\pi ^{2}T^{4}}{15}+\frac{\pi ^{2}T^{4}}{12}\left( 1-\psi
^{2}\right) ^{2}\text{.} 
\end{equation}
The one loop result dominates the effective potential for 
$T \gg \Lambda_{QCD}$ due to asymptotic freedom.
The first term is the standard black body result, obtained when $\psi = 1$.
Figure 1 shows $V$ as a function of $\psi$, normalized such that
$V(\psi = 1)=0$.
The use of the variable $\psi $ makes the $Z(2)$ symmetry of the potential
under $\psi \rightarrow -\psi \,$manifest. Note that the equilibrium value
of $\psi $ is $\pm 1$, corresponding to $L_{F}=\pm 1$; $\psi =0$,
corresponding to $L_{F}=0$, is a maximum of $V(\psi )$.

\vspace{1pt}Our picture of the quenching process is that the system is
initially in a state where $\psi \,$is equal to zero at some temperature
below $T_{d}$. When the system is quickly raised to a new temperature $%
T>T_{d}$, the system is still in the state with $\psi =0$. However, the
system is unstable, and must eventually find its way to either $\psi =+1\,$%
or $\psi =-1$. Because we quench into a region of the phase diagram
where $V^{\prime \prime }(\psi )<0$, the system will
decay to the equilibrium state via spinodal decomposition.

\section{Langevin Model}

\vspace{1pt}In order to study the dynamics of this transition, we use the effective
action of Bhattacharya combined
with Langevin dynamics.\cite{Bhattacharya} 
The effective action takes the form 
\begin{equation}
S_{eff}\left[ \psi \right] =\int d^{3}x\left[ \frac{\pi ^{2}T}{2g^{2}}\left(
\nabla \psi \right) ^{2}+\frac{\pi ^{2}T^{3}}{12}\left( 1-\psi ^{2}\right)
^{2}\right] \text{.} 
\end{equation}
The equilibrium distribution will be 
\begin{equation}
\exp \left[ -S_{eff}\left[ \psi \right] \right] \text{.} 
\end{equation}
We postulate Langevin dynamics of the form 
\begin{equation}
\frac{\partial \psi (x,\tau )}{\partial \tau }=-\Gamma \frac{\delta
S_{eff}\left[ \psi \right] }{\delta \psi (x,\tau )}+\eta (x,t) 
\end{equation}
where the white noise $\eta $ is normalized to 
\begin{equation}
\left\langle \eta (x,\tau )\eta (x^{\prime },\tau ^{\prime })\right\rangle
=2\Gamma \delta ^{3}(x-x^{\prime })\delta (\tau -\tau ^{\prime })\text{.} 
\end{equation}

Assuming translation-invariant initial conditions, the expectation value $%
\Psi (\tau )=\left\langle \psi (x,\tau )\right\rangle $ will evolve away
from the unstable value $\Psi=0$ approximately as

\begin{equation}
\frac{d \Psi }{d \tau }=\Gamma \frac{\pi ^{2}T^{3}}{3}\left(
1-\Psi ^{2}\right) \Psi 
\end{equation}
This gives initial exponential growth, which slows down as equilibrium is
approached.

The late-time relaxational behavior of $\psi $ is controlled by the Debye
screening mass $m_{D}$, given by $m_{D}^{2}=2g^{2}T^{2}/3$. Near
equilibrium, any effects of initial conditions decay exponentially as $\exp
\left[ -\frac{2\pi ^{2}\,\Gamma T}{g^{2}}(k^{2}+m_{D}^{2})\tau \right] $.
However, for early times, the initial conditions contribute to $\left\langle 
\widetilde{\psi }(k,\tau )\widetilde{\psi }(-k,\tau )\right\rangle $ a term
of the form 
\begin{equation}
\widetilde{\psi }(k,0)\widetilde{\psi }(-k,0)\exp \left[ -\frac{2\pi
^{2}\,\Gamma T}{g^{2}}(k^{2}-k_{c}^{2})\tau \right] 
\end{equation}
where $k_{c}^{2}$\vspace{1pt}$=g^{2}T^{2}/3=m_{D}^{2}/2$. Modes with $%
k<k_{c} $ are initially not damped but grow exponentially, with the $k=0$
mode growing the fastest. This is a characteristic feature of spinodal
decomposition with a non-conserved order parameter.

Because pure $SU(2)$ gauge theory has a second-order deconfining transition,
spinodal decomposition will occur after quenching to any temperature $%
T>T_{d} $. The situation is more subtle for $SU(3)$. 
The one-loop effective potential is unstable at $L_F = 0$.
For temperatures
sufficiently large that the one-loop effective potential for $L$ is
reliable, spinodal decomposition will occur. 
However, the first-order character of
the pure $SU(3)$ transition implies the existence of a metastable phase with 
$\left\langle L_{F}\right\rangle =0\,$for some range of temperature $%
T\gtrsim T_{d}$. This in turn implies
that nucleation followed by bubble growth is the important
mechanism for attaining equilibrium for some range of temperatures
just above $T_d$.

\section{Dynamical Quarks}

\vspace{1pt}When dynamical particles in the fundamental representation, e.g,
quarks, are included, the $Z(N)$ symmetry is explicitly broken. At low
temperatures, $L_{F}(x)\neq 0$. However, after a rapid quench to $T>T_{d}$, $%
L_{F}$ must still change to its new equilibrium value. This change may
involve spinodal decomposition or nucleation as well as relaxational
processes. 
The relevant potential for $SU(2)$ with $N_{f}$ massless
fermion flavors has the form 
\begin{equation}
V(\psi )=\frac{\pi ^{2}T^{4}}{12}\left( 1-\psi ^{2}\right) ^{2}+\frac{%
N_{f}\,\pi ^{2}T^{4}}{96}\left( 7+2\psi -\psi ^{2}\right) \left( 1-\psi
\right) ^{2}\text{.} 
\end{equation}
With $N_{f}=2$, $V^{^{\prime \prime }}(\psi )\,$\ is negative for $\psi <%
\frac{1}{3}$, as shown in figure 2, 
so spinodal decomposition will take place for a significant
range of initial conditions.\cite{Gross}\cite{Weiss}
Thus the initial value of $L_{F}$ in the confined phase
will determine whether spinodal decomposition occurs.
These considerations are independent of any static critical behavior, 
\textit{e.g.}, the existence of a deconfinement or chiral phase transition
for particular values of quark parameters.
Because $Z(N)$ symmetry breaking terms in the effective potential
vanish in the limit of infinite quark mass, spinodal decomposition
is to be expected for sufficiently large quark masses.

\vspace{1pt}

\section{Monte Carlo Results for $SU(2)$}

\vspace{1pt}

\vspace{1pt}We have tested these theoretical ideas with simulations of
rapidly quenched pure $SU(2)$ gauge theory. Lattices of size $32^{3}\times 4$
and $64^{3}\times 4$ were equilibrated at $\beta =2.0$; the deconfinement
transition for $N_{t}=4$ occurs 
at $\beta _{d}=2.2986\pm 0.0006$.\cite{Fingberg}
The coupling constant was increased
instantaneously to $\beta =3.0$, and the approach to equilibrium monitored
via the Polyakov loop and other observables. For simplicity, the heat bath
algorithm was used both for equilibration at low temperature
before the quench and for subsequent dynamical evolution after the
quench. While
this time evolution is of course not the true time evolution of the
non-equilibrium quantum field theory, features such as spinodal decomposition
which depend only on the equilibrium effective action will occur
with any local updating algorithm which converges to the equilibrium
distribution.
The abrupt change in $\beta $ is a potential cause of concern with this
procedure, since the lattice spacing, and hence the
physical volume, changes in all directions when $\beta $
is changed.
However, the large spatial sizes used should
mitigate this effect. 
It is also possible to maintain a constant spatial volume with
anisotropic lattice couplings. 
As an alternative, we have also taken $32^{3}\times 4$
sections from $32^{3}\times 10$ configurations equilibrated
at $\beta =3.0$, with similar
results. However, the initial configurations produced in this way suffer
from an obvious problem due to abrupt joining of the new boundaries in the
time direction.

In figure 3, we show the Polyakov loop expectation value versus Monte Carlo
time for one $64^{3}\times 4$ simulation. Comparison with 
a numerical integration of equation 3.5
shows qualitative agreement, although the equilibrium value of $\left\langle
L_{F}\right\rangle $ measured in simulations has a multiplicative
renormalization. Figure 4 shows the Fourier transform of the connected
Polyakov two point function $S(k,\tau )\,$for low values of the wave number
as a function of Monte Carlo time for the same simulation. Note the early
exponential rise in these modes, followed by a sharp disappearance as the
Polyakov loop reaches its equilibrium value, characteristic of spinodal
decomposition. Only the low momentum modes exhibit this growth; above $k_c$
no such growth occurs. Although the general behavior of $S(k,\tau )$ 
is the same for each run, many details are run dependent. In this
particular run, the $k/T = 0.68$ mode achieves a larger amplitude
than $k/T = 0.56$, which is atypical. In some runs, there is clear
evidence for mode-mode coupling, reflecting the nonlinearity of the system.
For each run,
we have estimated the rate of growth of each low-momentum mode by fitting
$log(S(k,\tau))$ to a straight line in $\tau$ for early times.
We can extract $k_c$ as the value where the growth rate is zero.
From equation 3.6, the growth rate of each line is proportional to $k_c^2-k^2$
for the linearized theory, but this may not represent the true
time evolution. In any case,
the growth rates measured in each run
are highly dependent on initial conditions.
In figure 5, we plot these growth rates versus $k^2/T^2$ for the same
run used in figure 4. 
The error bars are naive estimates of the error for each growth rate
for this particular run.
The x intercept
provides an estimate of $k_c^2$ for each run.
Using multiple $64^3$ runs at $\beta = 3.0$, we have estimated $k_c/T$ to be $1.09\pm 0.08$.
The principle errors in this estimate come
from the sensitivity of individual modes to initial conditions
and the discrete character of $k$ on the lattice.
We can compare this with lattice measurements of the Debye
screening length, assuming the one-loop relation
\begin{equation}
\frac{k_c}{T} = \frac{m_D(T)}{\sqrt{2}T}
\end{equation}
holds in general.
Using the results of Heller \textit{et al.}\cite{Heller} for $m_D(T)$, we obtain
$k_c/T =1.35(5)$. We consider this to be reasonable agreement,
given the many uncertainties involved.

Somewhat less sensitive to initial conditions is
the traditional observable $k^{\ast }$, which is defined to be 
\begin{equation}
k^{\ast }(\tau )=\frac{\int_{0}^{\infty }dk\,\,k\,S(k,\tau )}{%
\int_{0}^{\infty }dk\,\,S(k,\tau )}\text{.} 
\end{equation}
It is convenient to plot $T/k^{\ast }(\tau )\,$versus $\tau $, as in figure
6. Its early time behavior is predicted from equation 3.6 to be $1/k^{\ast
}(\tau )\sim \sqrt{\tau }$, typical of length scales in diffusive processes.
We do observe this behavior, but only for very early times; 
on $64^{3}\times 4\,$%
lattices, this occurs for $0\leq \tau \lesssim 120$.
For the run shown in figure 6, $T/k^{\ast }(\tau )$ is fit well by
$a\tau^p$ for $0\leq \tau \leq 120$, where $p = 0.54 \pm 0.03$.

\vspace{1pt}

\section{Conclusions}

As we have shown above,  theoretical and simulation results both indicate
the relevance of spinodal decomposition in the equilibration of a gluon
plasma after a rapid quench to high temperature. Since simulations and
theoretical analysis were both carried out in Euclidean space, the significance
of these results for experiment is not clear. However, spinodal
decomposition is a general phenomenon whenever a system is in an unstable
initial state, as determined by some local variable. Since we have
considered a rapid heating of a finite temperature gauge theory, it is
natural to ask what happens when rapid cooling takes place, as might occur
in the late stages of the expansion of a quark-gluon plasma, or in the early
universe. At low temperatures, we expect that $V(\psi )$ has a single
minimum and $V^{\prime \prime }(\psi )>0$ everywhere; this is required by $%
Z(N)\,$symmetry in the case of a pure gauge theory. Thus theory predicts the
absence of unstable modes, and that relaxational processes should dominate
the approach to equilibrium. We have performed simulations of such a cooling
process, in which $32^{3}\times 4$ lattice configurations equilibrated
at $\beta =3.0$ are suddenly cooled to $\beta =2.0$.
Examination of the data from these runs shows no sign of spinodal
decomposition.

It is interesting to contrast the phenomenon of spinodal decomposition in a
rapid heating with the formation of a disordered chiral condensate (DCC) in
a rapid cooling.\cite{DCC}
In both cases, exponentially growing low-momentum modes
occur when an unstable initial state must equlibrate to a final state
associated with a broken symmetry. In the case of a DCC, chiral symmetry is
the relevant symmetry. For massless quarks, the symmetry is spontaneously
broken at low temperatures. With sufficiently light quarks, the overall
structure of the potential survives, and the approach to equilibrium is
fundamentally the same as the case of massless quarks.
In the absence of fundamental representation particles such as quarks,
$Z(N)$ symmetry breaks spontaneously at high temperatures.
Provided there are sufficiently few light quarks,
spinodal decomposition still occurs at sufficiently high temperature.

\vspace{1pt}Just as the formation of a DCC may lead to enhanced production
of low-momentum pions, spinodal decomposition should lead to enhanced
production of low-momentum gluons in the early stages of plasma formation.
The characteristic scale for such a phenomenon would be $gT$. As the system
equilibrates, this enhancement of the small-$k$ part of the gluon
distribution will be eliminated. It is interesting to note that another,
perhaps related, mechanism for the creation of exponentially growing
low-momentum modes has been proposed; it has been
argued that such instabilities can significantly affect the temporal
evolution of the system.\cite{Mrowczynski} 

In contrast to the $SU(2)$ results presented here, the case of $SU(3)$
presents additional complications due to the presence of the confined phase
as a metastable state over some range of temperature in the deconfining
regime. Study of the $SU(3)\,$dynamics may be helpful in exploring the
limits of metastability. It would also be of interest to study via
simulation the effect of unquenched quarks on the dynamics. Because the
dynamic part of the simulation is much faster than creating equilibrated
field configurations, this could be done fairly easily if large equilibrated
unquenched lattice field configurations became available.

\section*{ACKNOWLEDGEMENTS}
We wish to thank the U.S. Department of Energy for financial support.

\pagebreak

\figure{ The effective potential $V(\psi)/T^4$ 
for $SU(2)$ with $N_f=0$ as a function of $\psi$.
\label{f1} }
\epsfysize=8in \epsfbox{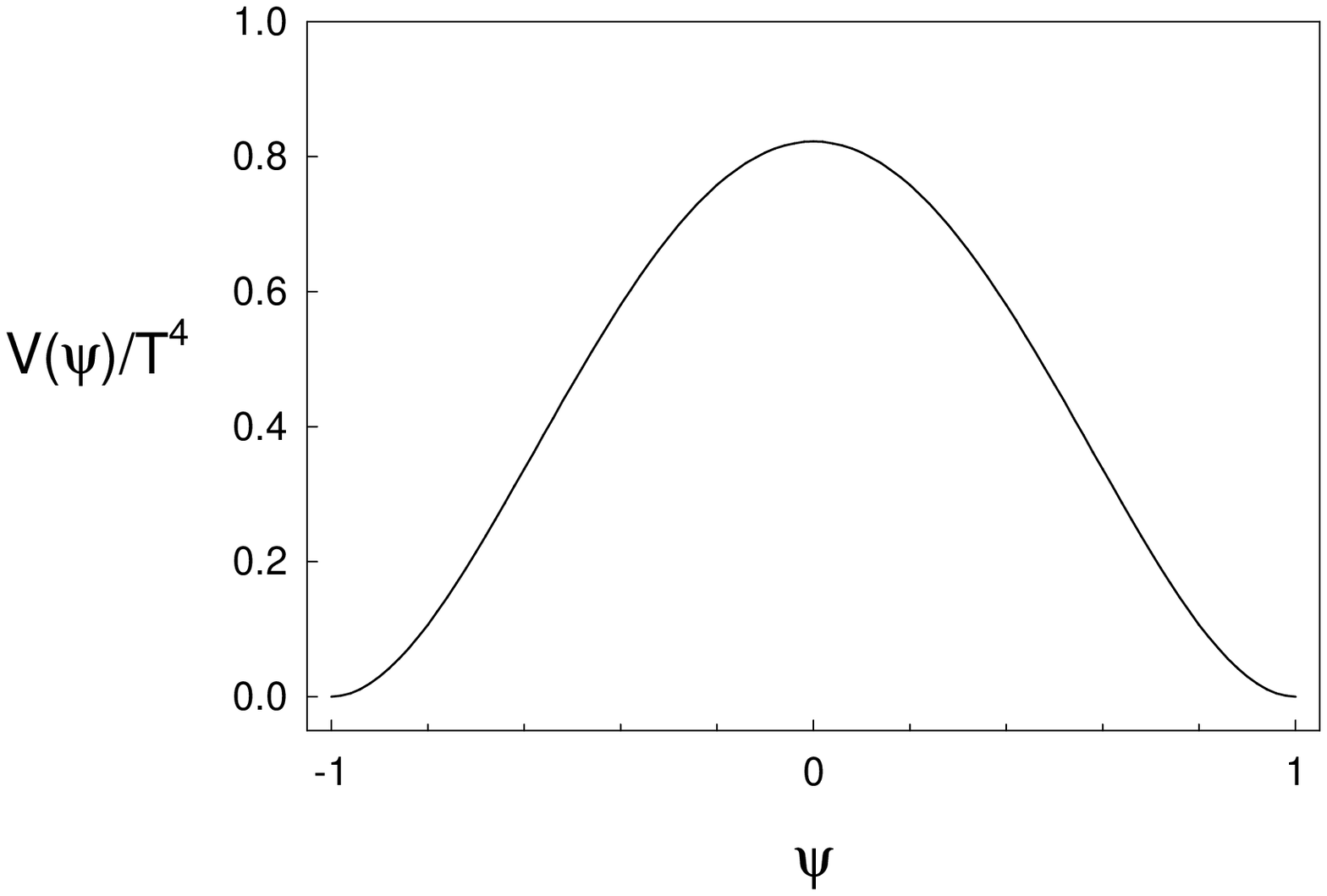}
\vspace{1.0in}
\pagebreak

\figure{ The effective potential $V(\psi)/T^4$ 
for $SU(2)$ with $N_f=2$ as a function of $\psi$.
\label{f2} }
\epsfysize=8in \epsfbox{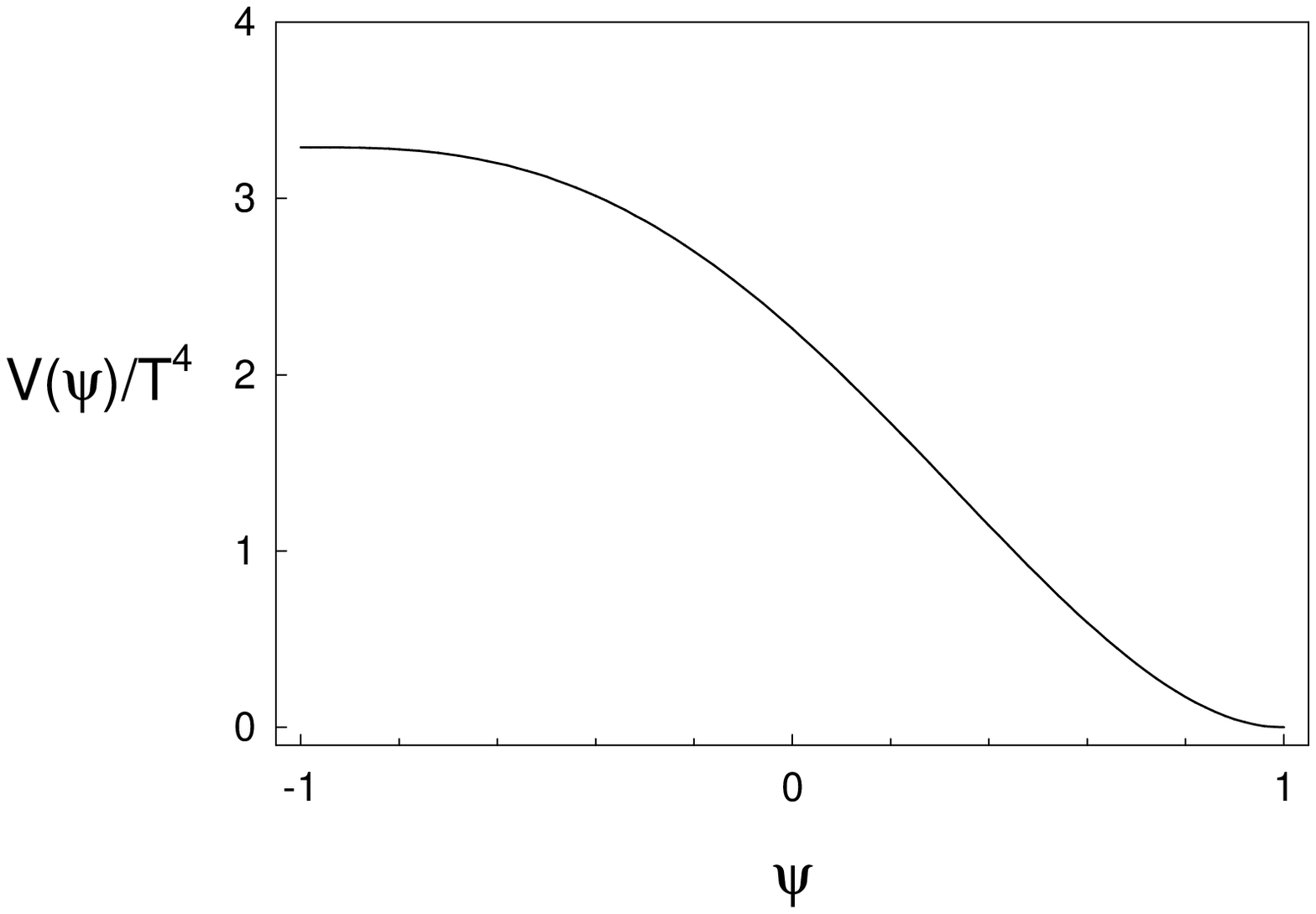}
\vspace{1.0in}
\pagebreak

\figure{ Polyakov loop as a function of Monte Carlo time.
\label{f3} }
\epsfysize=8in \epsfbox{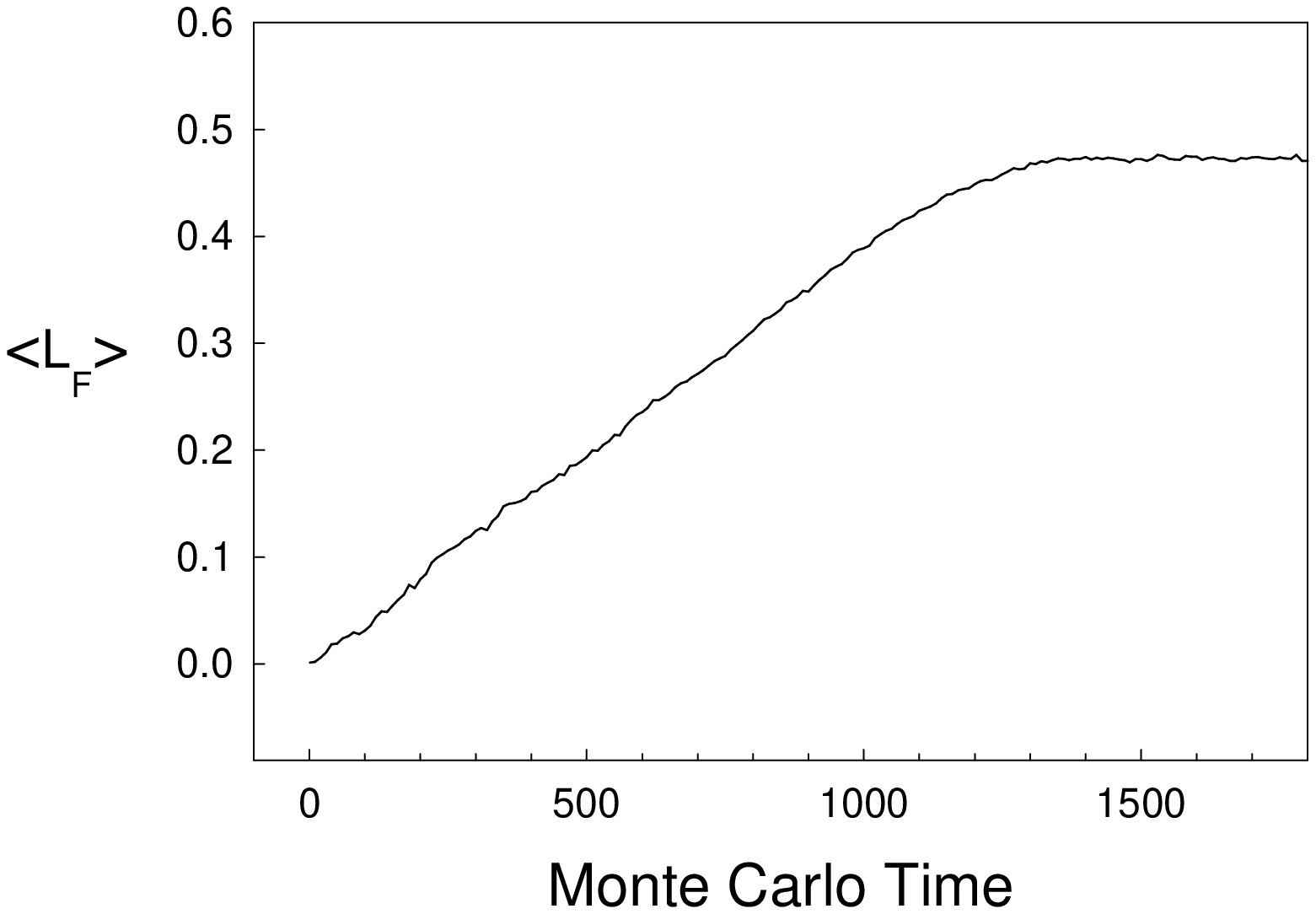}
\vspace{1.0in}
\pagebreak

\figure{ $S(k,\tau)$ versus Monte Carlo time.
\label{f4} }
\epsfysize=8in \epsfbox{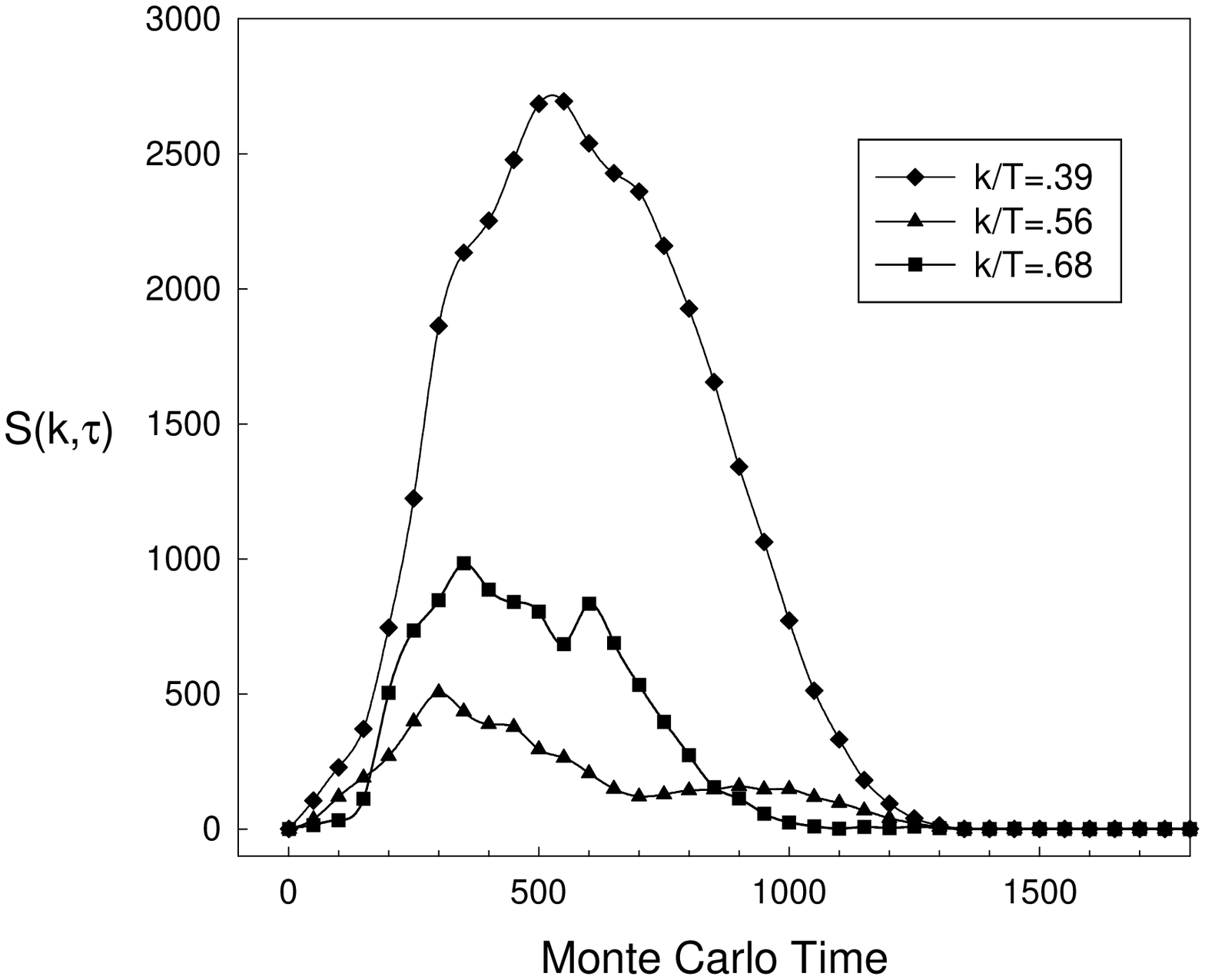}
\vspace{1.0in}
\pagebreak

\figure{ Growth rate of $S(k,\tau)$ versus $k^2/T^2$.
\label{f5} }
\epsfysize=8in \epsfbox{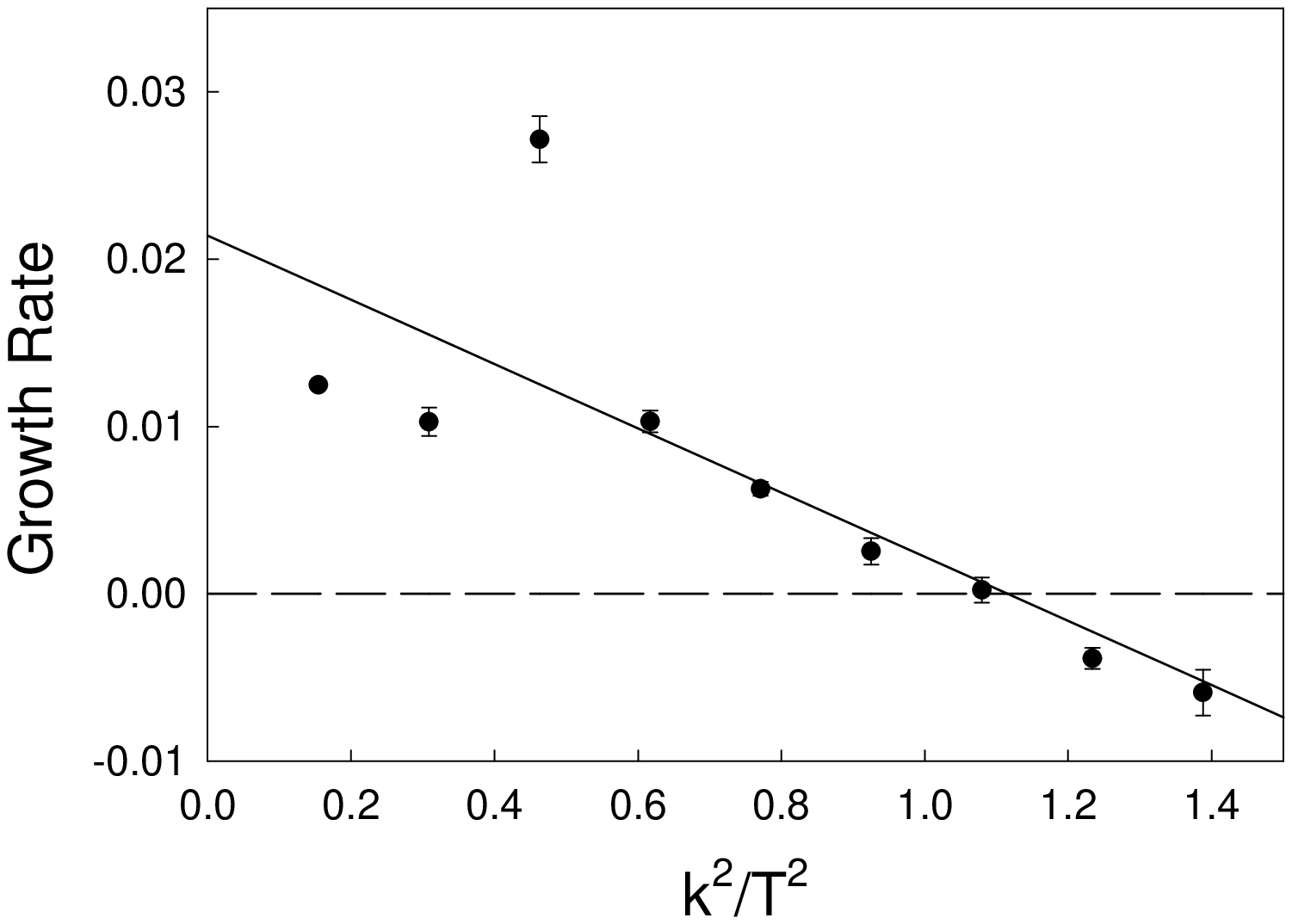}
\vspace{1.0in}
\pagebreak

\figure{ $T/k^*(\tau)$ versus Monte Carlo time.
\label{f6} }
\epsfysize=8in \epsfbox{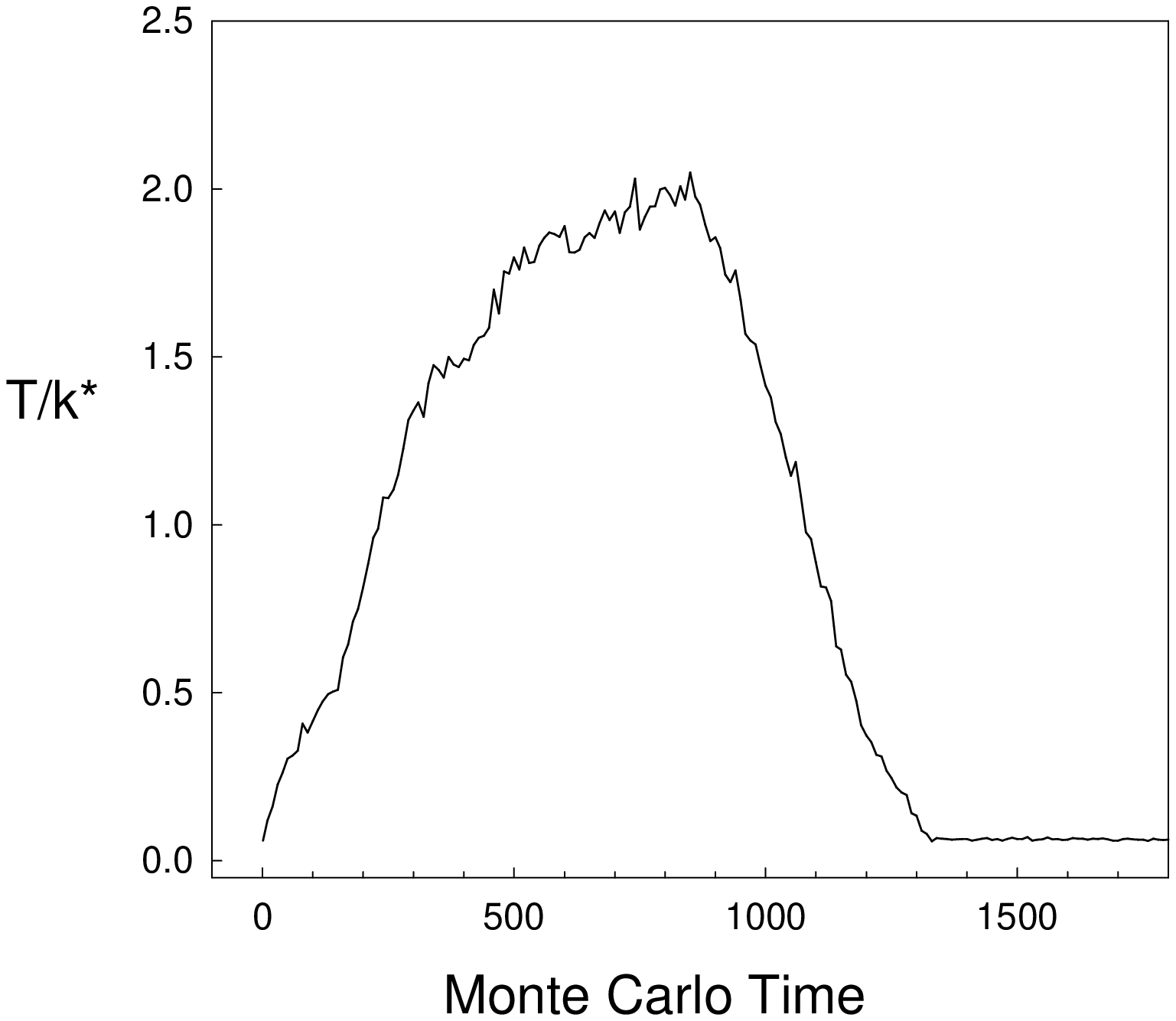}
\vspace{1.0in}
\pagebreak

\end{document}